\documentclass{mn2e}
\usepackage{graphicx}

\def \xor{$L_X/L_B$}
\def \lxlb{$L_X/L_B$}
\def \cxo{{\it Chandra X-Ray Observatory}}
\def \chandra{{\it Chandra}}
\def \ergs {\rm{erg\,s^{-1}}}
\def \kms{{km\,s$^{-1}$}}
\def \apjl{ApS}
\def \apjs{ApL}
\def \aaps{A\&AS}
\def \apj{ApJ}
\def \aj{AJ}
\def \mnras{MNRAS}
\def \araa{ARA\&A}
\def \z{\phantom{0}}

\title[X-ray emission properties of galaxies in Abell 3128]{X-ray emission properties of galaxies in Abell 3128}
\author[Russell J. Smith]
{Russell J. Smith\\Department of Physics, University of Waterloo, Waterloo, Ontario, Canada \ N2L 3G1}

\pagerange{\pageref{firstpage}--\pageref{lastpage}} \pubyear{2003}

\begin{document}

\label{firstpage}

\maketitle

\begin{abstract}

We use archival \cxo\ data to investigate X-ray emission from early-type
galaxies in the rich $z=0.06$ cluster Abell 3128. By combining the X-ray count-rates from an input list of 
optically-selected galaxies, we obtain a statistical detection of X-ray flux, 
unbiased by X-ray selection limits. Using 87 galaxies with reliable \chandra\ data, 
X-ray emission is detected for galaxies down to $M_B\approx-19.0$, 
with only an upper limit determined for galaxies at $M_B\approx-18.3$. 
The ratio of X-ray to optical luminosities is consistent with recent determinations of the
low-mass X-ray binary content of nearby elliptical galaxies. 
Taken individually, in contrast, we detect significant ($3\sigma$) flux for only six galaxies.
Of these, one is a foreground galaxy, while two are optically-faint galaxies with X-ray hardness 
ratios characteristic of active galactic nuclei. The remaining three detected galaxies are amongst 
the optically-brightest cluster members, and have softer X-ray spectra. Their X-ray flux is higher than 
that expected from X-ray binaries, by a factor 2--10; the excess suggests these galaxies have retained
their hot gaseous haloes. The source with the highest $L_X/L_B$ ratio is of unusual optical
morphology with prominent sharp-edged shells. 
Notwithstanding these few exceptions, the cluster population overall exhibits X-ray 
properties
consistent with their emission being dominated by X-ray binaries. 
We conclude that in rich cluster environments, interaction with the ambient
intra-cluster medium acts to strip most galaxies of their hot halo gas. 
\end{abstract}

\begin{keywords}
galaxies: clusters: individual: Abell 3128 --- 
galaxies: elliptical and lenticular, cD ---
galaxies: haloes --- 
X-rays: galaxies 
\end{keywords}

\section{Introduction}

Early-type galaxies emit X-rays through at least three mechanisms. 
Low-mass X-ray binaries (LMXBs) generate X-ray emission from an accretion disk
formed around a degenerate object in close binary systems, and are likely to be 
present in all such galaxies. Active galactic nuclei (AGN) are certainly harboured by 
some galaxies, generating X-rays from accretion onto a central super-massive black hole. 
Finally, some systems emit copious X-rays from a gravitationally-confined 
halo of hot ($\sim10^7$\,K) interstellar gas. 

The ratio of X-ray to optical luminosity, \xor, varies dramatically amongst early-type 
galaxies, spanning a range of up to two orders of magnitude between the 
so-called `X-ray bright' and `X-ray faint' classes (Canizares, Fabbiano \& Trinchieri 1987).
This great diversity is thought to arise mostly from the presence or absence of 
hot halo emission: while some galaxies have retained their interstellar medium, in others 
this gas has been removed, perhaps by ram-pressure stripping in galaxy clusters and groups (although
other  processes have been proposed). 
Many studies have searched for a dependence of the X-ray to optical luminosity ratio, \lxlb\,
on local environmental parameters, in an effort to establish the mechanism by which 
the haloes are lost or retained. To date however, no coherent picture has emerged, with some 
studies (e.g. White \& Sarazin 1991) reporting supressed X-ray emission in high-density environments, 
while others (e.g. Brown \& Bregman 2000) find increasing \lxlb\ with ambient density. 
In the case of galaxies which are dominant within their group or cluster, it is
particularly difficult to establish whether the X-ray emission arises from the galaxy, or 
rather from an intra-group medium (e.g. Helsdon et al. 2001). Analysing a homogenized 
compilation of ROSAT data, O'Sullivan Forbes \& Ponman (2001) concluded that \lxlb\ is not
{\it systematically} dependent on environment when AGN and group- (and cluster-) dominant galaxies are discounted. 

Early studies of X-ray emission from E/S0s were mostly restricted
to nearby galaxies in the field, in poor groups, and in the Virgo and Fornax clusters.
The densest environments, by contrast, can be probed only by studying rich galaxy clusters,
which lie at much greater distances. Such studies were undertaken using ROSAT/PSPC data by 
Dow \& White (1995) for the Coma cluster, and using ROSAT/HRI by Sakelliou \& Merrifield (1998)
for Abell 2634. Both of these works employed the technique of source `stacking' to obtain 
an unbiased statistical detection of X-ray flux from optically-selected galaxies. 

With the advent of high-resolution X-ray astronomy made possible by the \cxo, detailed studies of
point-sources in cluster fields have been reported by several authors (e.g. Sun \& Murray 2002; 
Martini et al. 2002). 
To our knowledge, however, there have been no published studies extending the 
source-stacking method to distant clusters using \chandra\ data. In this paper, 
we present just such an analysis, based on an archival observation 
of Abell 3128, a  rich, highly-substructured, cluster at $z=0.06$. 

\section{Observations and data reduction}

\subsection{Optical Imaging}

The present study is based on a catalogue of galaxies detected in optical imaging 
from the NOAO Fundamental Plane Survey (NFPS). The NFPS is a survey of $\sim$100 
rich low-redshift galaxy clusters, with principal science goals in the fields of 
large-scale structure (peculiar velocity measurements) and galaxy evolution, using
stellar population diagnostics. The NFPS provides mosaic imaging data
in the $B$ and $R$ bands, and follow-up intermediate dispersion spectroscopy for 30--90
$R<17$ red-sequence member galaxies per cluster. For an early summary of the NFPS, 
see Smith et al. (2000).

The optical imaging observations of Abell 3128 were obtained in September 1999, using the 8k Mosaic 
camera at the Blanco 4m Telescope of the Cerro Tololo Inter-American Observatory (CTIO). 
Two 90\,s images were obtained in $B$ and two 60\,s images in $R$. 
Reductions followed typical methods for large-format cameras, as implemented in the {\it mscred}
package of {\sc iraf}.
Objects were detected 
and galaxy catalogues compiled using SExtractor (Bertin \& Arnouts 1996). 
 Optical magnitudes quoted here are total (Kron-type) SExtractor 
magnitudes, corrected for $k$-dimming at the cluster redshift ($k_R=-0.06, k_B=-0.29$)
and uniform galactic extinction ($A_R=0.04, A_B=0.07$; Schlegel, Finkbeiner \& Davis 1998). 
The input list for X-ray flux measurements consists of all galaxies satisfying 
$R<18.0$ and $B<20.0$.
Over this range the cluster exhibits a very prominent 
red-sequence, indicating the large fraction of passive E/S0 galaxies. The optical 
catalogue contains $\sim$250 galaxies, but of these only 104 are covered by the \chandra\ 
observation discussed below. 

\begin{figure*}
\centering
\includegraphics[width=170mm]{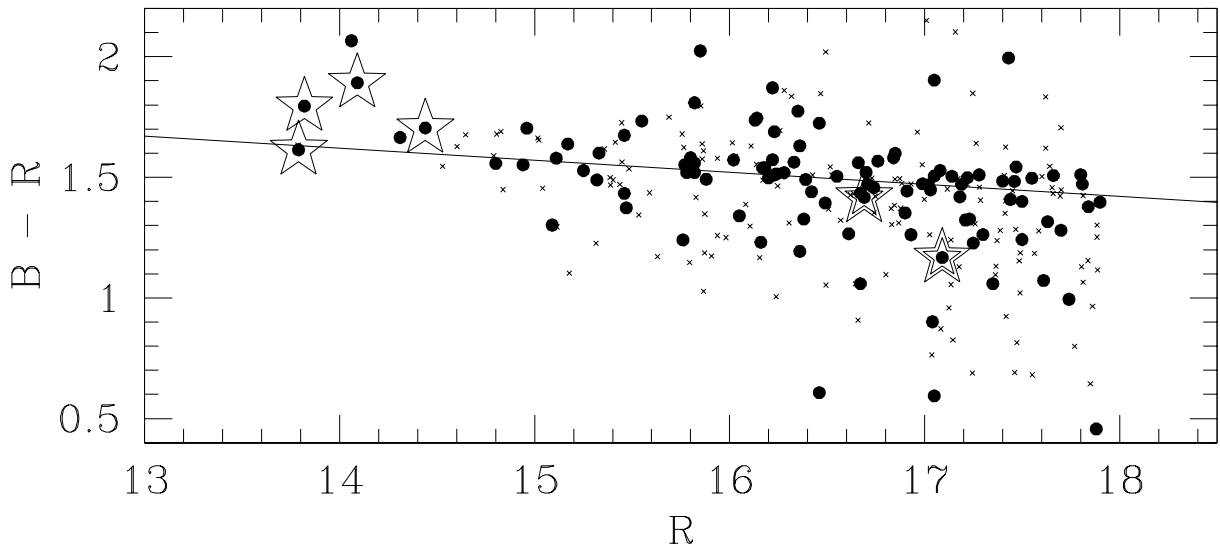}
\vskip -87.5mm
\caption{Optical $(B-R)$ colour--magnitude relation for NFPS sample galaxies in Abell 3128. 
Filled circles mark galaxies with overlapping X-ray data, while small crosses represent galaxies 
outside of the \chandra\ field of view, or with unreliable X-ray data (see text). 
Large stars mark galaxies with significant X-ray flux 
detections at 0.5--2.0\,keV (the brightest of these is actually a foreground galaxy). 
Those with `double' stars are also detected in the 2.0--10.0\,keV band, and are probably AGN.}
\label{cmr}
\end{figure*} 

\subsection{X-ray data}

The \chandra\ data derive from a 20\,ks observation in the ACIS-I configuration
(PI J. Rose). The observation placed the cluster A3128 at the aimpoint of the 
ACIS-I array. The diffuse emission has been analysed by Rose et al. (2002), who 
discuss the `double-peaked' morphology of the cluster in X-rays. 

The data were reduced using {\sc ciao 2.2} (and calibration data in {\sc caldb 2.15}), 
following standard processing `threads'. No background
flares were apparent in the light curve, so no time filters were applied. Since high 
angular resolution is critical for reducing contamination from the cluster background 
and from neighbouring sources, we consider only the four CCDs of the ACIS-I array, 
covering an area $15\times15$\,arcmin. 

X-ray counts were  measured at positions defined by the optical galaxy catalogue, 
after correction for an astrometric offset of $0.7\pm0.1$\,arcsec 
established using a handful of unambiguous bright sources. 
At each galaxy position, counts were summmed over the energy range 0.5-2.0\,keV, in a
circular `source' aperture, and also in a concentric `background' annulus. 
The source aperture radius was varied with location in the ACIS array, so as to include 99\% of 
the local point-spread function 
flux, computed at an enery of 1\,keV. The counts were weighted according to the exposure map,
likewise computed for an energy of 1\,keV, yielding a common flux system over the array. 
To convert count rates to fluxes in the 0.5--2.0\,keV band, we assumed a thermal bremsstrahlung
spectral model, with a temperature of 8.1\,keV. This model is appropriate for sources
dominated by LMXBs (Sarazin, Irwin \& Bregman 2001). 
In converting to X-ray 
luminosity, we adopt a cosmological model with $(\Omega_m, \Omega_\Lambda, h) = (0.3, 0.7, 0.7)$,
yielding a luminosity distance of 273\,Mpc for z=0.060. All sources are assumed to lie at the cluster distance. 
Luminosities are corrected for galactic neutral hydrogen absorption with column density 
$n_{\rm H}=1.47\times10^{20}$\,cm$^{-2}$
(Dickey \& Lockman 1990). 

In order to improve the reliability of the photometry, we rejected sources lying in regions of
high background intensity. This primarily affects galaxies in the two bright X-ray clumps at the
north--east and south--west of the field (Rose et al. 2002). Furthermore, we rejected measurements
in which the centroid of the background counts was significantly ($>$3$\sigma$) displaced from the 
aperture centre; this criterion is especially valuable in removing measurements where the 
background estimate is compromised by a bright point source. After removing these sources, the matched
X-ray-optical catalogue contains flux measurements for 87 galaxies. 

Strictly, these flux measurements are valid only for point sources. If the X-ray emission 
is distributed similarly to the stellar light (as may be approximately the case for LMXBs), then 
this assumption slightly underestimates the total flux. 
Based on measured half-light radii, the typical aperture correction is
$\Delta\log{}L_X\approx0.04$, while $\sim$6 galaxies would require corrections of
$\Delta\log{}L_X=0.15-0.22$. 

The above procedure was also applied to measure counts in a harder energy range
(2.0-10.0\,keV). For this case, we applied exposure map and point-spread 
function estimates appropriate for a 6\,keV mono-energetic source. 

\section{Results}

\subsection{Individually-detected sources}

Figure~\ref{cmr} shows the $B-R$ colour--magnitude relation for the input catalogue 
galaxies, highlighting those objects with significant ($>3\sigma$) X-ray detections. 
The properties of these sources are also summarized in Table~\ref{sources}.

The six detected sources fall into two apparent groups in colour-magnitude space. 
The first group contains four of the brightest galaxies on (and redwards of) the 
E/S0 sequence. These objects are detected with high significance in the 
0.5--2.0\,keV band, but only marginally in the hard (2.0-10.0\,keV) band. In fact, Source 1
is a foreground object with $z=0.039$ (Katgert et al. 1998). The other three optically-bright
sources are confirmed cluster members. 
The X-ray emission from Sources 2 and 4 is twice that expected from LMXBs (see below). 
Source 3 outshines the LMXB predictions by a factor $\sim$10. Interestingly, this galaxy 
lies close to the North--East peak of the cluster emission (Rose et al. 2002), but 
not coincident with it. The NFPS spectrum of this source suggests a quiescent giant elliptical, but
the morphology is unusual: the galaxy has an extended sharp-edged outer shell or
envelope. It is tempting, given these cD-like characteristics, to identify Source 3 as the 
former dominant galaxy of an infalling subcluster. However, the heliocentric
recession velocity of this galaxy, from the NFPS spectrum, is 17241\,\kms, which is not consistent with 
the redshift-space group suggested by Rose et al. to be associated with 
the North--East X-ray peak ($cz=$18600--19400\,\kms). 

The remaining two X-ray sources (5 and 6) are optically much fainter, and lie on the blue side 
of the CMR. The redshifts of these sources are not known. In addition to their soft emission, these
galaxies were also detected at high significance ($>3\sigma$) in the hard bandpass. Given the 
hard-to-soft flux ratios
and high luminosities of these sources, we consider it likely that X-ray emission in
these galaxies arises primarily from AGN activity. 
If these are cluster members, the AGN fraction amongst the sample galaxies is $\sim$2.5\%, subject
of course to very large uncertainties. This may be compared with recent \chandra\ estimates
of $\sim5\%$ for Abell 2104 (Martini et al. 2002) and $\sim$12\% for Abell 1367 (Sun \& Murray 2002). 
Selecting by optical spectroscopy yields an AGN fraction of $\sim$1\% in clusters 
(Dressler et al. 1999), but misses low-luminosity or heavily obscured sources.

\subsection{Stacked-source results}

\begin{figure*}
\centering
\includegraphics[width=170mm]{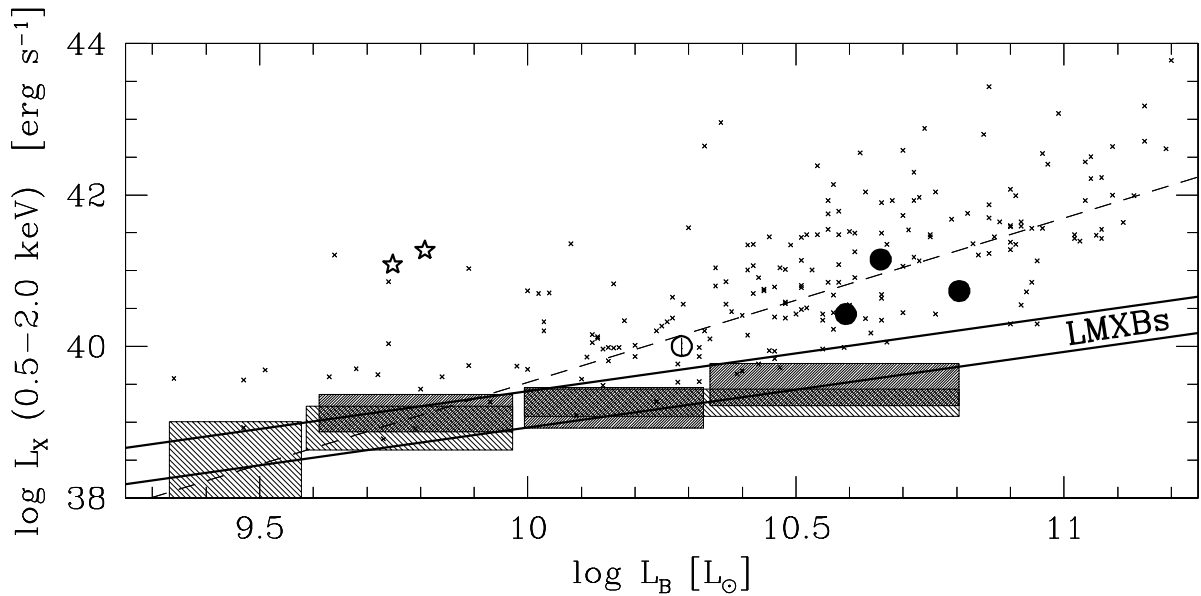}
\vskip -85mm
\caption{X-ray luminosities from stacked sources, computed over four bins in B-band luminosity. 
The large symbols show the four galaxies with significant detections: filled circles are confirmed cluster
members, the open circle is a known foreground galaxy (luminosities computed for the true redshift). 
The two open stars indicate the probable AGN, with luminosities computed as if they were at the cluster redshift. 
The shaded boxes show the mean X-ray 
luminosity in each bin, with the vertical extent indicating the error 
(only an upper limit is obtained for the faintest bin). The horizontal extent of the box indicates the
luminosity range of galaxies in the bin. The dark-shaded boxes show results for confirmed cluster 
members only, with `aperture' corrections applied. 
Two solid lines of unit slope indicate the expected trend if LMXBs dominate the
X-ray emission. Finally, the small points and dashed line represent the detections of O'Sullivan et al. (2001), together
with their bias-corrected fit.}
\label{sakmer}
\end{figure*}

\begin{table*}
\begin{center}
\caption{Abell 3128 galaxies detected at $>3\sigma$ in the 0.5--2.0\,keV band.}
\label{sources}
\begin{tabular}{ccccccl}
\hline
ID & R.A. & Dec 			& $R$   & $B-R$  & $L_X$ [0.5--2.0\,keV] &  Notes \\ 
  & \multicolumn{2}{c}{[J2000]} 	& [mag] & [mag]  &  [$10^{40}\,\ergs$]  \\
\hline
1& 03:29:53.125 & --52:30:54.70 & 13.79 & 1.614  & \z0.8$\pm$0.2 & Foreground object z=0.039\\
\hline
2& 03:30:38.429 & --52:37:09.64 & 13.82 & 1.795  & \z4.1$\pm$0.9 & Normal E/S0\\  
3& 03:30:51.032 & --52:30:31.32 & 14.09 & 1.892  &  10.7$\pm$1.5 & E/S0 with shells\\ 
4& 03:30:13.684 & --52:37:30.04 & 14.44 & 1.704  & \z2.0$\pm$0.6 & Normal E/S0 \\ 
\hline
5& 03:30:17.322 & --52:34:08.72 & 16.69 & 1.417  & 14.0$\pm$1.6   & Hard spectrum : AGN?\\ 
6& 03:29:41.460 & --52:29:36.01 & 17.09 & 1.167  & \z9.1$\pm$1.3  & Hard spectrum : AGN?\\ 
\hline
\end{tabular}
~\\
{\scriptsize Note: Luminosities computed assuming the cluster distance except for source $\#1$}
\end{center}
\end{table*}

For the purpose of combining fluxes for galaxies grouped by luminosity, we exclude the two
galaxies with strong hard X-ray fluxes (whether AGN or not, these clearly do not belong 
to the general population of cluster galaxies). In addition, of course, the criteria
already mentioned are used to reject galaxies in regions of high diffuse X-ray emission, 
and those with asymmetric backgrounds. 

The mean stacked-source fluxes and their errors are computed in four bins of B-band luminosity, 
as shown in Figure~\ref{sakmer}. Only an upper limit is obtained for the faintest luminosity bin. 
Also shown in this figure are estimates of the LMXB contribution, scaling from the LMXB population
observed in nearby ellipticals. The upper line is based on the results of Sarazin et al. (2001), who
estimated the total flux from LMXBs in the nearby X-ray faint elliptical NGC 4697, yielding 
$L_{\rm LMXB} (0.3-10\,$keV$) / L_B = 8.1\times10^{29}\,$erg\,s$^{-1}$\,$L_\odot^{-1}$. Similar values
have been obtained for NGC 720 (Jeltema et al. 2003) and NGC 1553 (Blanton, Sarazin \& Irwin 2001). 
In the absence of a wider study employing consistent source detection and analysis conventions, it is
difficult to generalize these results or estimate the intrinsic distribution of 
LMXB-to-optical luminosity ratio. From an unpublished compilation, White, Sarazin \& Kulkarni (2002) 
claim a factor of four variation in this ratio; there is evidence that many LMXBs are created in globular 
clusters, so that their contribution may depend on cluster frequency (White et al. 2002; Kundu, Maccarone \& Zepf 2002).
To reflect the uncertain range, we simply set the lower line in Figure~\ref{sakmer} to a factor of three lower 
than that of NGC 4697. In calculating the LMXB predictions for the bandpass adopted here, we model
the LMXB emission using a 8.1\,keV thermal bremsstrahlung spectrum  (the best-fitting model of Sarazin et al.). 

The small points in Figure~\ref{sakmer} represent X-ray detected galaxies in the ROSAT compilation
of O'Sullivan et al. (2001). Their fit (dashed line) is based on a survival-statistics analysis
of these detections, together with upper limits not shown here. 
There is a striking difference between the ROSAT sample and our data for Abell 3128: at the 
bright end, our stacked-source measurements lie a factor of $\sim$20 fainter in X-rays than 
the O'Sullivan et al. sample. The steep slope of the ROSAT sample is not reproduced here; rather, 
the slope is consistent with direct proportionality with the optical luminosity. Finally, the 
overall zero-point of the $L_X-L_B$ relationship is, for our sample, fully consistent with the
contribution expected from LMXBs alone. These results argue strongly that, at least in the very 
richest environments, most cluster galaxies are indeed deprived of their hot X-ray haloes.

We have assumed in these calculations that all NFPS galaxies lie at the cluster distance. 
Where cluster membership can be confirmed using NFPS spectra and literature data (via NED), 
similar results are obtained, differing only in the last bin where no redshifts are available. 
Figure~\ref{sakmer} shows, as dark shaded boxes, the results for confirmed cluster members. 
In addition, these latter points are corrected for source-extension based on the measured 
half-light radius, as discussed above. It is seen that these corrections are very small 
in comparison to the size of the observed effect.
To test the assumption that early-type galaxies dominate the sample, a two-dimensional profile-fitting
code (GIM2D: see Simard et al. 2002) 
was used to identify $\sim$20 galaxies with small bulge-to-disk ratios. 
Removing these galaxies has little effect on the results. 

\section{Discussion}

The principal result of this {\it Letter} is that the X-ray emission from cluster members can be 
accounted for by the LMXB population alone. 
With the probable exception of the $\sim3$ galaxies noted in the previous section, 
galaxies in this very rich cluster do not show signs of emission from haloes of hot gas. 
This result can be interpreted as evidence for a strong environmental influence on the X-ray properties, 
with gas haloes being removed from cluster galaxies through, for example, ram-pressure stripping by the 
intra-cluster medium (White \& Sarazin 1991). 

Our result is qualitatively similar to that of Sakelliou \& Merrifield (1998), who applied a 
similar method to high-spatial resolution ROSAT/HRI data for Abell 2634, concluding that 
LMXBs account for the entirety of the emission. Similarly, an earlier application of the 
source-stacking technique to Coma, by Dow \& White (1995), found X-ray luminosities compatible 
with LMXBs, at least for the fainter galaxies in their sample. 
(The brightest galaxies are coincident with enhancements in the diffuse cluster gas, from which
they cannot be unambiguously separated in their ROSAT/PSPC observation.)

This apparently consistent picture from studies using source-stacking contrasts starkly with 
the inconclusive results of `compiled' galaxy samples.
For example,  White \& Sarazin (1991) claimed that X-ray bright 
ellipticals have fewer neighbours than their X-ray faint cousins (a result which
argues in the same sense as ours), while  O'Sullivan et al. (2001) have concluded that no 
coherent environmental mechanism drives the enormous range in $L_X/L_B$. Brown \& Bregman 
(2000) claim a contrary result, with $L_X/L_B$ increasing with density. 

It seems probable that this confused picture arises partly from difficulties in
disentangling galaxy emission from that of a surrounding group or cluster (Helsdon et al. 2001).
The `compiled' samples moreover contain very few galaxies in extremely rich (Coma-like) clusters, and
those which are included are typically the brightest cluster members. The source-stacking method
applied to clusters has the advantage of probing `typical' early-type galaxies in the very environments 
where stripping is likely to be most efficient.

\section{Conclusions}

Our X-ray source-stacking analysis of of E/S0 galaxies in Abell 3128 demonstrates that the cluster
population is dominated by `X-ray faint' objects, with typical X-ray to optical flux-ratio $\sim$20
times smaller than typical for field/group samples. Similar results have previously been obtained for Coma 
and Abell 2634. The natural explanation is in terms of a greater tendency for stripping of gas 
haloes in the densest environments. 

Excluding probable AGN, only one detected cluster member is as X-ray luminous as a 
typical field/group elliptical. It has an unusual optical morphology with an outer shell or
envelope. 
Two further bright galaxies have emission in somewhat in excess of the LMXB predictions.  

The data presented here thus yield a mixed picture in which most galaxies are stripped, but
just a few apparently retain a hot halo. The distribution and properties of the latter
should provide insights into the mechanisms which drive this dichotomy, 
but more cases are clearly required. We are currently applying the source-stacking analysis to a 
larger number of clusters, which will provide 
a sample large enough for meaningful comparisons between galaxies of morphological classes and in 
clusters of differing richness. 

\section*{ACKNOWLEDGMENTS}

I gratefully acknowledge the generous assignment of NOAO observing resources to the
NFPS programme; I thank my collaborators for allowing me to present these results in advance of the NFPS publications.
(CTIO) is operated by the Association of Universities for Research in
Astronomy, Inc. under a cooperative agreement with the National Science
Foundation. Jim Rose, as referee, provided several helpful comments. 
This research has made use of the NASA/IPAC Extragalactic Database
(NED) which is operated by the Jet Propulsion Laboratory, California
Institute of Technology, under contract with the National Aeronautics
and Space Administration.
{\sc Iraf} is distributed by the National Optical Astronomy Observatories
which is operated by the Association of Universities for Research in
Astronomy, Inc. under contract with the National Science Foundation. 

{}

\label{lastpage}
\end{document}